\documentclass[conference]{IEEEtran}
\IEEEoverridecommandlockouts
\usepackage{verbatim}
\usepackage[utf8x]{inputenc}
\usepackage{epstopdf}
\usepackage[pdftex]{graphicx}
\usepackage{tabularx}
\usepackage{booktabs}
\usepackage[table]{xcolor}
\usepackage{booktabs}
\usepackage{ctable}
\usepackage[cmex10]{amsmath}
\usepackage{multicol}
\usepackage{algorithm}
\usepackage{algpseudocode}
\usepackage{amssymb}
\usepackage{cite}
\usepackage{array}
\usepackage{caption}
\usepackage{subcaption}
\def\BibTeX{{\rm B\kern-.05em{\sc i\kern-.025em b}\kern-.08em
		T\kern-.1667em\lower.7ex\hbox{E}\kern-.125emX}}

\usepackage[letterpaper, top=.75in, bottom=.95in,left=0.625in, right=0.625in]{geometry}

\begin{document}
\bstctlcite{IEEEexample:BSTcontrol}
\title{Distributed Wideband Spatio-Spectral Sensing for Unlicensed Massive IoT Communications} 
\author{\IEEEauthorblockN{Ghaith Hattab and  Danijela Cabric}
	\IEEEauthorblockA{Department of Electrical and Computer Engineering\\
		University of California, Los Angeles\\
		Email: ghattab@ucla.edu, danijela@ee.ucla.edu}
\thanks{This work has been supported by the National Science Foundation under grants 1527026 and 1149981.}
}

\maketitle

\begin{abstract}
In this paper, we propose a dynamic spectrum sensing-based architecture to provide connectivity for a massive number of Internet-of-things (IoT) objects over the unlicensed spectrum. Specifically, the architecture relies on deploying sensing access points (SAPs), e.g., small cells with sensing capabilities, that aim to (i) identify a large number of narrowband channels in a wideband spectrum, as many massive IoT applications have low-rate requirements, and (ii) aggressively reuse the unlicensed channels at the SAPs' locations as IoT devices typically transmit at low power, occupying a small spatial footprint. Instead of enforcing each SAP to sense the entire spectrum, we develop a sensing assignment scheduler that ensures each one senses a subset of the spectrum. We then develop a distributed spatio-spectral cooperative sensing algorithm that enables each SAP to have local information about the occupancy of the entire spectrum. We present numerical simulations to validate the effectiveness of the proposed system in the presence of WiFi access points (APs). It is shown that the proposed system outperforms non-cooperative and centralized schemes in terms of reliably identifying more available spatio-spectral blocks with a lower misdetection of transmitting WiFi APs.  
\end{abstract}


\section{Introduction}
The development of fifth-generation new radio (5G-NR) has created use cases that transcend the constant interest in mobile broadband communications. Indeed, one of the new use cases of 5G-NR is supporting massive Internet-of-things (mIoT) communications, where IoT objects are connected to the Internet, enabling large-scale applications such as smart cities, public safety, and agriculture \cite{3GPP2015a,ITU2015}. To this end, new cellular categories are introduced such as narrowband IoT (NB-IoT) \cite{Rico-AlvarinoYavuz2016}, which aim to reduce device complexity and connect more devices over narrowband channels. However, due to the high congestion in the licensed spectrum and the high cost of owning it, a growing number of IoT-based networks have centered around the use of the unlicensed spectrum  \cite{Raza2017,MulteFire2017}.

Unlicensed-based IoT solutions, e.g., LoRa \cite{Raza2017}, primarily access fixed narrow spectrum bands, and rely on spread spectrum techniques for coexistence, but these do not scale well when the number of devices is very high. Further, they mainly use the sub-1 GHz ISM band \cite{Raza2017}, which is of bandwidth 26MHz, as it has favorable propagation conditions. In contrary, the MulteFire specification \cite{MulteFire2017}, a standard that enables a stand-alone unlicensed access of cellular networks, relies on small cells to support industrial and private IoT networks over the 5GHz spectrum. To reach the massive scale, we propose an architecture where sensing access point (SAPs), or small cells, sense a wideband spectrum with fine spectral resolution to find many narrowband channels and fine spatial resolution to reuse these channels across the SAPs. Such architecture can complement access protocols used by MutleFire networks.

The contributions of this paper are twofold. First, to limit the sensing burden on each SAP, we formulate an integer program, where the objective is to assign each SAP a subset of the wideband spectrum to sense. The sensing assignment problem is combinatorial with high complexity in dense networks, and thus we develop a heuristic low-complexity algorithm to perform the sensing assignments. Sensing assignment has been studied before in different context in \cite{Mochaourab2015,Lai2015,Zhang2015}. For instance, in \cite{Mochaourab2015,Lai2015}, the assignment is done such that each channel is sensed by one device, whereas in this work we require each channel to be sensed by multiple SAPs for reliable cooperative sensing. In \cite{Zhang2015}, the assignment aims to maximize the rate of secondary users, and thus it requires these users to sense all channels before making the assignment. In this paper, the assignment is done prior to sensing, where we aim to ensure that each SAP, sensing a specific subset of channels, is surrounded by SAPs sensing other subsets. The second contribution is the development of a distributed sensing algorithm, where each SAP senses its assigned subset of channels, shares and collects measurements from nearby SAPs, and processes the collected data to infer the spectrum occupancy across all channels. Different from the distributed sensing proposed in \cite{SobronVelez2015}, each SAP may arrive at a different decision as the occupancy of a channel vary over space. In addition, we use the combine-then-adapt diffusion algorithm \cite{ChenSayed2015} and propose a novel update of the algorithm's weights to quickly diffuse information about the wideband spectrum at each SAP. We validate the effectiveness of the proposed system via Monte Carlo simulations, where we test the sensing performance in the presence of WiFi access points (APs). Results show that the proposed system finds more available channels at each SAP, henceforth denoted as \emph{spatio-spectral blocks}, with a lower misdetection in comparison with non-cooperative and centralized cooperative schemes. We also simulate a massive IoT application, showing that the proposed system helps serve significantly more IoT devices compared to existing schemes.



\section{System Model and Proposed Architecture}\label{sec:model}
We consider a dense deployment of APs, denoted by the set $\mathcal{K}=\{1,2,\cdots,K\}$, where $K\gg1$. All of them are assumed to be connected to a core network. Furthermore, the $k$-th AP is connected to those in vicinity, which are denoted by the set $\mathcal{N}_k$. In this paper, we assume that any AP within distance $R$ from the $k$-th AP belongs to $\mathcal{N}_k$. 

The network is assumed to provide Internet connectivity over an unlicensed wideband spectrum of bandwidth $B$. Since the majority of mIoT applications have low-rate requirements, we assume that the spectrum is divided into narrowband channels, denoted by $\mathcal{M}=\{1,2,\cdots,M\}$, where $M=\lfloor\frac{B}{b}\rfloor$ and $b\ll B$ is the bandwidth of each channel. Any part of the spectrum can be also occupied by other incumbent networks. 

An example of such model is a cellular network that consists of $K$ small cells or femto base stations (BSs), e.g., a MulteFire deployment over the unlicensed spectrum \cite{MulteFire2017}. Each small cell is connected to the core network via the NG interface, i.e., the standard interface connecting BSs to the core, and neighboring BSs can communicate with each other using the Xn interface, i.e., the standard interface that connects 5G-NR BSs. The network may use MulteFire-based specifications or 5G-NR Stand-alone Unlicensed Access, where a wideband spectrum at $5$GHz and bandwidth $B\approx 500$MHz can be used for access. Using the NB-IoT operation, the channel bandwidth is $b=180$KHz \cite{Rico-AlvarinoYavuz2016}, and hence $M\approx 2800$. The incumbent transmitters in this case are primarily WiFi networks. We note further that the Federal Communications Commission (FCC) opened an inquiry on the use of the 5.9GHz-7.1GHz spectrum for 5G-NR \cite{FCC2017b}, i.e., in this case $B=1.2$GHz  and $M\approx 6666$. 

\begin{figure}[t!]
	\center
	\includegraphics[width=3.25in]{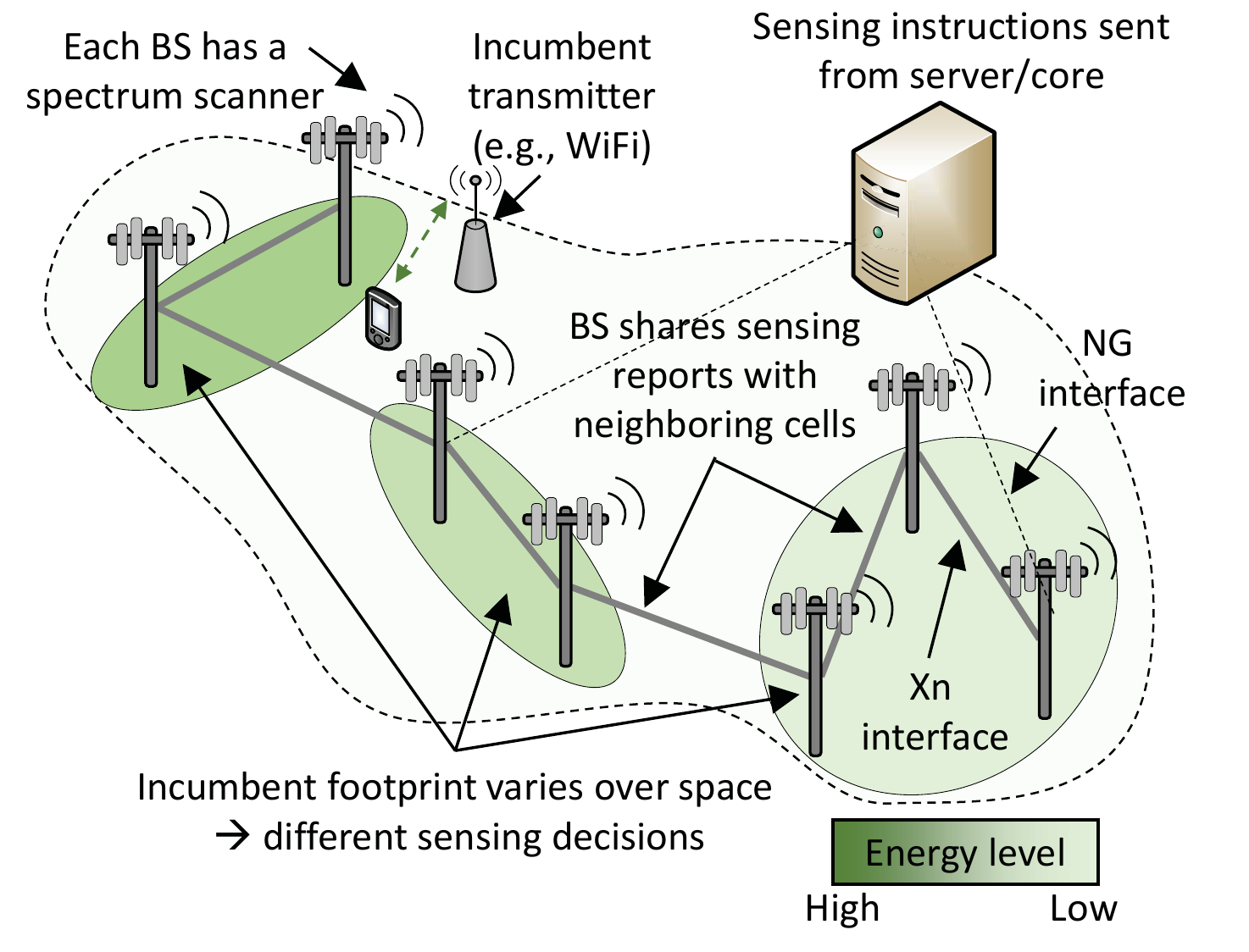}
	\caption{An illustration of the DSA-based architecture.}
	\label{fig:architecture}
\end{figure}

The proposed DSA-based architecture envisions equipping each AP with a spectrum scanner, henceforth denoted as sensing access points (SAPs), as shown in Fig. \ref{fig:architecture}. To connect a large number of IoT devices, it is critical to identify many narrowband channels in a wideband spectrum. However, the computational complexity to sense a wideband spectrum of order $B$ at a fine resolution to identify channels, each of bandwidth $b\ll B$ can limit the deployment of spectrum scanners at a large scale. Thus, the proposed architecture includes a sensing assignment scheduler that aims to reduce the sensing burden on each SAP. Specifically, the objective is to find an assignment across SAPs such that the $k$-th one senses $p_k\ll M$ channels, yet each $m$-th channel in $\mathcal{M}$ is sensed by $q_m>1$ SAPs for reliable sensing decisions. Using the cellular network example with NB-IoT operation, the SAP may sense only a 20MHz of 500MHz spectrum, i.e., $p_k=111\ll M$. Once the channel assignment is completed, each SAP locally senses the assigned channels, processes the sensing data, and shares it with neighboring SAPs.\footnote{In this work, we focus on identifying narrowband channels for massive IoT applications. For scheduling and access, we can follow the same protocols used in MulteFire \cite{MulteFire2017}. Future research directions may include jointly optimizing sensing and access, and the dissemination of channel occupancy information.} In this paper, we consider energy-based sensing, where we denote  the $m$-th channel energy measured at the $k$-th SAP by $Y_{k,m}$. Examples of systems using energy-based sensing are those that rely on listen-before-talk protocols, e.g., MulteFire small cells (or unlicensed cellular networks) use the energy detector, where the BS initiates a random back-off procedure if the measured energy level over a channel exceeds $-72$dBm \cite{MulteFire2017} (or $-62$dBm \cite{MukherjeeLarsson2016}). 
We model the received power spectrum over the $m$-th channel as \cite{Laghate2017}
\begin{equation}
\label{eq:MultibandDetectionProblem}
Y_{k,m}= V_{k,m}	+ \sum_{i} S_{i,k,m},
\end{equation}
where $V_{k,m}$ is the noise power and $S_{i,k,m}$ is the received signal power from an $i$-th incumbent transmitter over the $m$-th channel. The received signal power takes into account small-scale fading, large-scale fading, and shadowing. Furthermore, since the sensing algorithm will rely on SAPs sharing information with their neighbors, we assume an initial phase where the $k$-th SAP measures the power of a reference signal broadcasted by the $j$-th SAP, which is denoted by $\hat P_{k,j}\forall j\in\mathcal{N}_k$. Such measurement can be used to assess the quality of the sensing reports received from neighboring SAPs. A summary of the main parameters are given in Table \ref{tab:parameters}.

\begin{table}[!t]
	\caption{Main parameters}
	\label{tab:parameters}
	\centering
	\begin{tabular}{|c|l|}
		\hline
		Symbol   	&  Description\\\hline
		$\mathcal{K}$ and $K$ 	& The set of SAPs and $K=|\mathcal{K}|$\\
		$\mathcal{N}_k$  	& The neighborhood of the $k$-th SAP\\		
		$\mathcal{M}$ and $M$ 	& The set of channels and $M=|\mathcal{M}|$\\
		$B$		& Bandwidth of the wideband spectrum (Hz)\\
		$b$		& Bandwidth of the channel (Hz)\\
		$p_k$		& Number of channels to be sensed by the $k$-th SAP\\
		$q_m$		& Number of SAPs sensing channel $m$\\
		$Y_{k,m}$		& Measured power of $m$-th channel at the $k$-th SAP\\
		$\hat P_{k,j}$		& Reference signal power of $j$-th SAP at the $k$-th SAP\\
		\hline
	\end{tabular}
\end{table}

\section{The Sensing Assignment Scheduler}\label{sec:scheduler}
Consider the $k$-th SAP sensing only $\mathcal{M}_k\subset\mathcal{M}$ channels, which implies that by the end of the sensing period, the SAP will lack information about the remaining $(\mathcal{M}_k)^c=\mathcal{M}\setminus \mathcal{M}_k$ channels. To extract information about a channel in $(\mathcal{M}_k)^c$, the $k$-th SAP needs to collect sensing reports from neighboring SAPs that sensed that channel. However, it is unreliable to collect such reports from distant SAPs, as the spatial footprints of incumbents vary over space. Let $c_{j,k,m}$ denote the cost of the $k$-th SAP using the $j$-th SAP sensing report of the $m$-th channel, e.g, the cost can be the quality of the received power of reference signals broadcasted by the neighboring SAPs. Let such costs be collected in a matrix $\mathbf{C}_m\in\mathbb{R}^{K\times K}$. Further, let $\mathbf{X}\in\mathbb{Z}^{K\times M}$ be the assignment matrix, i.e., the $(k,m)$-th entry $x_{k,m}=1$ if the $k$-th SAP is assigned to sense the $m$-th channel, and 0 otherwise. Then, the general assignment problem is formulated as an integer program as follows.\footnote{An underlying assumption here is $\sum_{m=1}^M q_m= \sum_{k=1}^K p_k$.}
\begin{equation}
\label{eq:IPOriginal}
\begin{array}{cl}
\underset{\mathbf{X}}{\text{{minimize}}}
&~~ \underset{m\in\{1,2\cdots,M\}}{\operatorname*{max}} \sum_{j=1}^K \sum_{k=1}^K c_{j,k,m} x_{k,m}\\\
\text{subject to}     
&~~\sum_{k=1}^K x_{k,m}=q_m, ~~\forall m\\
&~~\sum_{m=1}^M x_{k,m}=p_k, ~~\forall k\\
&~~x_{k,m}\in\{0,1\}\\
\end{array}
\end{equation}
By minimizing the maximum cost across all channels, the $k$-th SAP can improve the reliability of the sensing reports collected regarding all channels in $(\mathcal{M}_k)^c$. Such framework is a generalization to the bottleneck assignment problem, which is NP-hard \cite{Martello1995}.  We remark a more practical framework may add the constraint that all channels to be sensed by the same SAP are consecutive ones. Due to the difficulty of this problem, we consider a simpler framework as follows.

First, we assume all SAPs sense the same number of channels, i.e., $p_k=p\forall k$. In addition, the spectrum is divided into $L=\lfloor\frac{B}{p\cdot b}\rfloor$ subsets, i.e., each SAP senses a single subset of $p$ consecutive channels, and each subset is sensed by $\tilde q_l$ SAPs. Thus, let $\tilde{\mathbf{X}}\in\mathbb{Z}^{K\times L}$ denote the sensing assignment matrix, with $\tilde x_{k,l}=1$ when the $k$-th SAP is assigned the $l$-th subset. Let $\tilde{c}_{j,k,l}$ be the cost of the $k$-th SAP using the $j$-th SAP sensing report of the $l$-th subset, which we can assume to be the maximum of the costs of the reports of the channels belonging to the $l$-th subset. Then, we consider the following simpler integer program. 
\begin{equation}
\label{eq:IPSimplers}
 \begin{array}{cl}
 \underset{\tilde{\mathbf{X}}}{\text{{minimize}}}
	&~~ \underset{l\in\{1,2\cdots,L\}}{\operatorname*{max}} \sum_{j=1}^K \sum_{k=1}^K \tilde c_{j,k,l} \tilde x_{k,l}\\\
 \text{subject to}     
 &~~\tilde{\mathbf{X}}^T\mathbf{1}_K=\tilde{\mathbf{q}}, \\
 &~~\tilde{\mathbf{X}}\mathbf{1}_L=\mathbf{1}_K, \\
 &~~\tilde x_{k,l}\in\{0,1\},\\
\end{array}
\end{equation}
where the $(k,l)$-th entry of $\tilde{\mathbf{X}}$ is $\tilde x_{k,l}$,  $\mathbf{1}_{l}\in\mathbb{R}^l$ is the one vector, and $\tilde{\mathbf{q}}=[\tilde q_1,\tilde q_2,\cdots,\tilde q_L]^T$. The optimization problem in (\ref{eq:IPSimplers}) has a lower complexity than $(\ref{eq:IPOriginal})$ as $L\ll M$, and it is practical as each SAP will sense a single block of $p$ channels instead of $p$ not-necessarily consecutive narrowband channels. However, it is still a combinatorial problem with high computational complexity when $K\gg1$. In what follows, we present a low-complexity heuristic assignment scheduler.

\subsection{A heuristic sensing assignment scheduler}
Consider the $l$-subset of channels. This subset must be sensed by $\tilde q_l$ SAPs such that the maximum cost of reports collected by SAPs not sensing this subset is minimized. Intuitively, collecting reports from very far SAPs should be discouraged as the spatial footprint of an incumbent varies over space. In other words, the set of SAPs sensing the $l$-th subset cannot be clustered in a given area, but rather they should be spread out over the region to ensure that any SAP not sensing the subset has a nearby SAP sensing it. This motivates us to first divide all SAPs into $\tilde q_l$ clusters $\mathcal{C}_{i,l}\subset{\mathcal{K}}\forall i=1,2,\cdots,\tilde q_l$. From each cluster $\mathcal{C}_{i,l}$, we pick one SAP that minimizes the worst cost, i.e., we solve the following problem for each cluster
\begin{equation}
\label{eq:IPCluster}
\begin{array}{cl}
\underset{\{\tilde x_{k,l}\in\mathcal{C}_{i,l}\}}{\text{{minimize}}}
&~~   \underset{j\in \mathcal{C}_{i,l}}{\operatorname*{max}}\sum_{k\in \mathcal{C}_{i,l}} \tilde c_{j,k,l} \tilde x_{k,l}\\\
\text{subject to}     
&~~\sum_{k\in\mathcal{C}_{i,l}} \tilde x_{k,l}=1,\\
&~~\tilde x_{k,l}\in\{0,1\}\\
\end{array}
\end{equation}
The optimal solution, in fact, is $\tilde x_{k,l}^\star=\{1|k=\operatorname*{argmin}_e  \sum_{j\in\mathcal{C}_{i,l}}\tilde c_{j,e,l}\}$. After this iteration, there remains $K-\tilde q_l$ SAPs. Thus, we pick another subset of channels, say $u$, and then cluster the remaining SAPs into $\tilde q_u$ clusters, and solve (\ref{eq:IPCluster}) for each cluster, repeating the process until all subsets are completed. Note that the subsets that are picked earlier in the procedure will have lower total cost as there are more SAPs to pick from. To combat this, we repeat the whole process multiple times, randomizing the order of picked subsets in each time. Then, we pick the one with the lowest maximum total cost. The proposed algorithm is summarized in Alg. \ref{alg:scheduler}. We remark that in this paper, we use the $k$-means clustering algorithm to find $\mathcal{C}_{i,l}$ for the $l$-th subset due to its low complexity.  

\begin{algorithm}[!t]
	\small
	\caption{Proposed spectrum assignment scheduler}\label{alg:scheduler}
	\begin{algorithmic}[1]
		\Procedure{Assignment}{$\mathcal{K}, \tilde{\mathbf{C}}$}
		\For{$n=1\longrightarrow N$}
		\State \textbf{Set} $\tilde{\mathcal{K}}=\mathcal{K}$		
		\State \textbf{Permutate} $\boldsymbol{l}=[1,2,\cdots,L]$
		\For{$u=1\longrightarrow L$}
		\State \textbf{Partition} $\tilde{\mathcal{K}}$ into $\tilde q_{l(u)}$ clusters $\{\mathcal{C}_{i,l(u)}\}_{i=1}^{\tilde q_{l(u)}}$
		\For{$i=1\longrightarrow \tilde q_{l(u)}$}
		\State \textbf{Solve} (\ref{eq:IPCluster}) to compute $\tilde x^\star_{k,l(u)}\forall k\in\mathcal{C}_{i,l(u)}$
		\EndFor
		\State \textbf{Update} $\tilde{\mathcal{K}}\rightarrow \tilde{\mathcal{K}}\setminus\{k|\tilde x_{k,l(u)}^\star=1\forall \mathcal{C}_{i,l(u)},i=1,\cdots,\tilde q_{l(u)}\}$ 
		\EndFor
		\State \textbf{Store} $\tilde{\mathbf{X}}^\star_n$ and its corresponding objective value $Z_n$
		\EndFor		
		\State \textbf{Return} $\tilde{\mathbf{X}}^\star_{n^\star}$, where $n^\star = \operatorname{argmin}_n Z_n$
		\EndProcedure
	\end{algorithmic}
\end{algorithm}

\subsection{Numerical Validation}
\begin{figure}[t!]
	\center
	\includegraphics[width=3.25in]{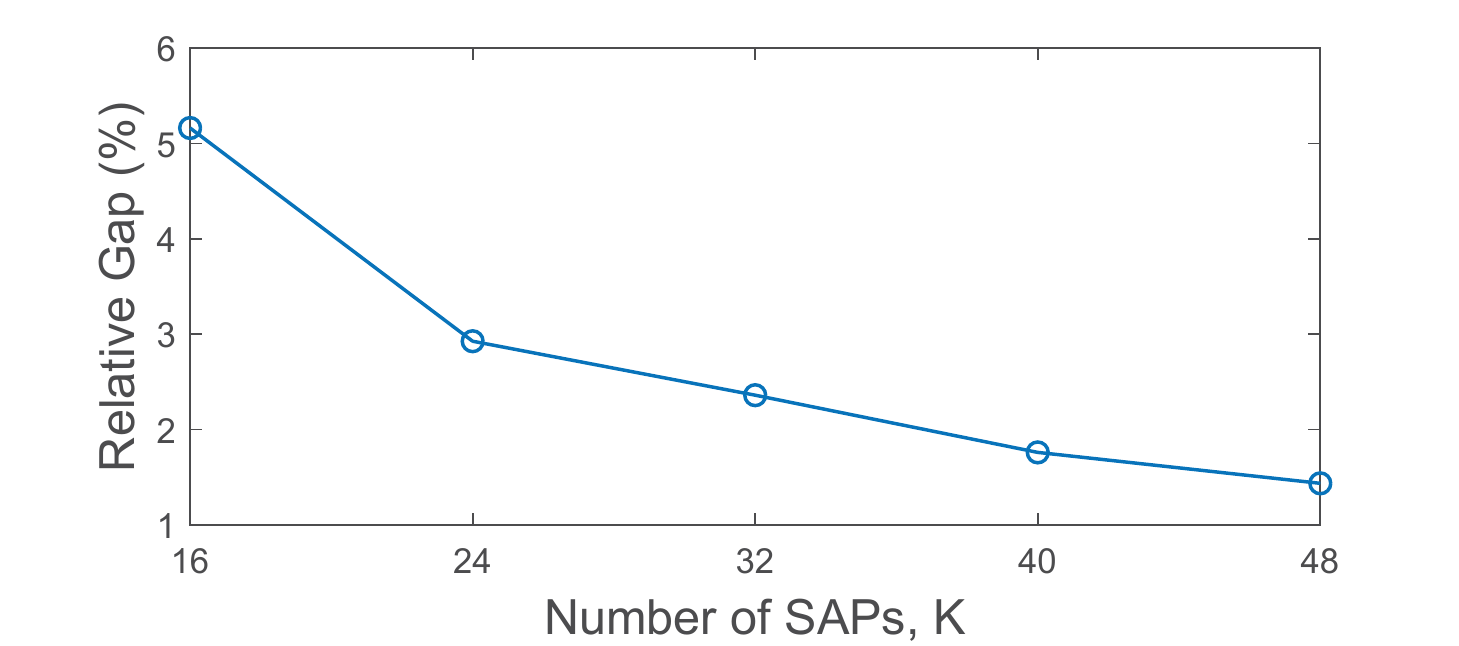}
	\caption{Relative gap between the objective function values of (\ref{eq:IPSimplers}) using the integer program and Alg. \ref{alg:scheduler}.}
	\label{fig:schedulerComparison}
	\vspace{-0.1in}
\end{figure}

We compare the solution of the integer program in (\ref{eq:IPSimplers}) to the solution of Alg. \ref{alg:scheduler}, where (\ref{eq:IPSimplers})  is solved using MOSEK 8.0 solver with CVX in MATLAB. Due to the high complexity of solving the latter problem, we only compare the two solutions using relatively small number of SAPs. We run 50 different realizations, where in each one we randomly deploy SAPs on an area of $2\times2\text{km}^2$, and assume uniform costs, i.e., $\tilde c_{j,k,l}\sim\mathcal{U}(0,1000)$ for comparison.  Here, we assume the number of subsets is $L=4$, and $\mathbf{q}=q\mathbf{1}$, and hence the number of SAPs is $K=q\cdot L$. Fig. \ref{fig:schedulerComparison} shows the gap between the objective function in (\ref{eq:IPSimplers}) when evaluated at the optimal solution of the integer program and the solution obtained by the proposed algorithm for different number of SAPs. It is observed that the algorithm performs very well relative to the integer program, and the gap reduces for higher density of SAPs, as higher density provides more flexibility of sensing assignments. 

\section{Diffusion-based Distributed Sensing}\label{sec:sensing}
Using raw energy estimates to make decisions about the activity in a channel can be unreliable, particularly in the presence of fading. A common way to tackle this is to combine multiple energy estimates collected from several spectrum scanners, yet the output of the combiner can still fluctuate abruptly from one sensing window to another \cite{SobronVelez2015}. Therefore, it is desired to use and share a more robust statistic for reliable cooperative sensing. To this end, an adaptive least-mean-squares (LMS) filter is proposed in \cite{SobronVelez2015} to soften the energy measurements. Specifically, the spectrum scanner aims to minimize the cost function $J(w_m)= \mathbb{E}[(d_{k,m}-w_m Y_{k,m})^2]$ by optimizing $w_m$, where it is desired to have $d_{k,m}= \mathbb{E}[Y_{k,m}]$ to reduce the variance of the different energy measurements. Since the receiver does not have prior information about $\mathbb{E}[Y_{k,m}]$, $d_{k,m}$ is estimated in an online manner. 

While it is shown in  \cite{SobronVelez2015} that the proposed approach significantly improves the detection performance in comparison with the energy detector, the approach requires all SAPs to have the same $w_m$ to optimize. In other words, a cooperative procedure in this case implies that all SAPs will aim to find the optimal $w_m$ that minimizes a global cost function, forcing each SAP to have the same decision on the occupancy of the $m$-th channel since $w_m$ is used as the test statistic. Since we aim to aggressively reuse channels over space, we need a fine-resolution frequency-space map of the spectrum. To this end, we propose to define a SAP-specific cost function, i.e.,  $J_k(w_{k,m})= \mathbb{E}[(d_{k,m}-w_{k,m} Y_{k,m})^2]$.
Although each SAP has a different $w_{k,m}$ to optimize, nearby SAPs that sense the $m$-th channel should have similar optimal solutions due to the spatial correlation. Hence, cooperation among neighboring SAPs can improve the spectrum sensing reliability parallel to capturing the spatial variations of the incumbent's footprint. Thus, the SAPs cooperate to minimize $J_0(\boldsymbol{w}) = \sum_{k=1}^K J_k(\boldsymbol{w})$.

We aim to solve this global optimization distributively using diffusion-based distributed algorithms, which are known to be superior to LMS-based algorithms \cite{ChenSayed2015}. In this algorithm, each SAP has a vector of weights $\boldsymbol{w}_{k}=[w_{k,1},w_{k,2},\cdots,w_{k,M}]^T$ to compute. Different from existing diffusion-based algorithms \cite{ChenSayed2015}, the $k$-th SAP only optimizes the vector entries corresponding to its assigned subset of channels, instead of optimizing all entries. The other entries will be  computed using the measurement reports of the SAPs that sense different subsets. 

The algorithm is centered around two main stages: the combination stage and the adaptation stage. Specifically, each SAP shares its estimated $w_{k,m,i-1}$ (the subscript $i$ denotes the iteration number) with its neighbors for the combination stage, where we propose the following combination policy
\begin{equation}
\label{eq:combineStep}
\psi_{k,m,i-1} = \left\{
\begin{array}{ll}
 \sum_{j\in\mathcal{N}_{k}} \alpha_{jk,m,i} w_{j,m,i-1}, & m\in \mathcal{M}_k\\
 \sum_{j\in\mathcal{N}_{k}} \beta_{jk,i} w_{j,m,i-1}, &m\notin \mathcal{M}_k\\
\end{array}\right.
\end{equation}
where $\{\alpha_{jk,m,i},\beta_{jk,i}\}$ are the combination weights, and they satisfy $\sum_{j\in\mathcal{N}_{k}} \alpha_{jk,m,i} =1$, $\sum_{j\in\mathcal{N}_{k}} \beta_{jk,i}=1$, $\alpha_{jk,m,i}\geq0$, and $\beta_{jk,i}\geq0$. In the adaptation stage, the following update formulation is used
\begin{equation}
\begin{aligned}
\label{eq:weight}
w_{k,m,i} &= \psi_{k,m,i-1}\\
&+ \operatorname{{I}}_{(m\in\mathcal{M}_k)}\mu_{k} Y_{k,m,i}\left[d_{k,m,i}-Y_{k,m,i}\psi_{k,m,i-1}\right],
\end{aligned}
\end{equation}
where $\operatorname{{I}}_{(\cdot)}$ is the indicator function, $\mu_{k}$ is a constant step-size, and $d_{k,m,i}$ is a first-order filter to a approximate $\mathbb{E}[Y_{k,m}]$ \cite{SobronVelez2015}. That is, we have $d_{k,m,i}   = \zeta d_{k,m,i-1} + (1-\zeta) Y_{k,m,i}$, where  $\zeta$ is a scalar close but less than one. 

The combination policy is central to this distributed sensing algorithm. In particular, we aim to perform an online clustering of SAPs with similar measurements, that runs in parallel, by adapting the combination weights $\alpha_{jk,m,i}$. If the $k$-th SAP is sensing the $m$-th channel, it assigns the following weight to the channel report sent from the neighboring $j$-th SAP \cite{ChenSayed2015}
\begin{equation}
\label{eq:combinationweight_alpha}
\alpha_{jk,m,i} = \frac{(w_{k,m,i-1}+\mu_{k} \gamma_{k,m,i}-w_{j,m,i-1})^{-2}}{\sum_{j\in\mathcal{N}_{k}} (w_{k,m,i-1}+\mu_{k} \gamma_{k,m,i}-w_{j,m,i-1})^{-2}},
\end{equation}
where $\gamma_{k,m,i}=(d_{k,m,i}-Y_{k,m,i}w_{k,m,i-1})Y_{k,m,i}$. This weighting mechanism looks at the similarities between the estimated $w_{k,m,i-1}$ and $w_{l,m,i-1}$, where higher weight is given when these two values are closer to each other. Hence, as the algorithm progresses, cooperating SAPs become clustered based on the similarities of their optimal solutions. If the $m$-th channel is not sensed by the $k$-th SAP, then it will collect the reports from the neighboring SAPs that sense this channel and combine them using the following weight	
\begin{equation}
\label{eq:combinationweight_beta}
\beta_{jk} = \frac{\hat P_{k,j}}{\sum_{j\in\mathcal{N}_{k}\setminus k} \hat P_{k,j}},
\end{equation}
where we have  dropped the subscript $i$ as only one reference signal received power, per SAP, is used. 

After $N$ iterations, each $k$-th SAP will have $\boldsymbol{w}_{k,N}=[w_{k,1,N},w_{k,2,N}\cdots, w_{k,M,N}]^T$, which will be compared with a threshold vector $\boldsymbol \lambda_{k}=[\lambda_{k,1},\lambda_{k,2}\cdots, \lambda_{k,M}]^T$ to make a decision on each channel. The threshold vector can be computed for instance by feeding the algorithm with samples of known energy, e.g., $-62$dBm \cite{MukherjeeLarsson2016}, and using the output as a threshold for future samples. A summary of the proposed algorithm is given in Alg. \ref{alg:diffusion}. Note that this algorithm has low sensing complexity since each SAP scans a subset of the spectrum and shares the sensing data only with nearby SAPs, i.e., the communication overhead for sharing the weights per SAP does not scale with the total number of SAPS but rather with the number of neighbors of each one.

\begin{algorithm}[!t]
	\small
	\caption{Proposed distributed sensing algorithm implemented by the $k$-th SAP}\label{alg:diffusion}
	\begin{algorithmic}[1]
		\Procedure{Diffusion}{$\mu_{k},\mathcal{N}_{k}$,$\zeta$,$\hat P_{k,j}$}
		\For{$i=1\longrightarrow N$}
		\State \textbf{Measure} $Y_{k,m,i}$
		\State \textbf{Estimate} $d_{k,m,i}   = \zeta d_{k,m,i-1} + (1-\zeta) Y_{k,m,i}$
		\State \textbf{Compute} $\gamma_{k,m,i}=(d_{k,m,i}-Y_{k,m,i}w_{k,m,i-1})Y_{k,m,i}$ 
		\State \textbf{Compute weights} $\alpha_{jk,m,i}$ using (\ref{eq:combinationweight_alpha}) and $\beta_{jk}$ using (\ref{eq:combinationweight_beta}) 
		\State \textbf{Combine} $\psi_{k,m,i-1}$ using (\ref{eq:combineStep}) and \textbf{Adapt}  $w_{k,m,i}$ using (\ref{eq:weight})
		\EndFor
		\State \textbf{Compare} $\boldsymbol{w}_{k,N} \gtrless\boldsymbol\lambda_{k}$
		\EndProcedure     
	\end{algorithmic}
\end{algorithm}

\section{Simulation Results}\label{sec:simulations}
In this section, we compare the performance of the proposed architecture with different schemes via Monte Carlo simulations. We first present the results over a small-scale network to visualize the different decisions made by SAPs, and then evaluate the schemes in a large-scale mIoT application.

In the first set-up, we consider 100 SAPs in a grid deployment and an inter-site distance of 200m. The neighborhood set of the $k$-th SAP includes all SAPs within 200m. We assume the number of channels is four, each of bandwidth 20MHz. We then randomly deploy 50 WiFi access points, where each one randomly picks one of the channels. We assume all WiFi APs transmit at a fixed power of $30$dBm. For the channel model, we consider the 3GPP NR-UMi model \cite{3GPP2017d}, which is suitable for dense urban areas. We assume that the spectrum is centered around 5.43GHz. All SAPs and WiFi APs are assumed to be at a height of 10m. We run 1000 realizations, where the channels used by WiFi APs are randomized and the propagation losses are varied from one realization to another due to fading and the log-normal shadowing in  the 3GPP NR-UMi model \cite{3GPP2017d}. We then consider the following schemes:
\begin{itemize}
\item \textbf{Genie}: A system that has access to all true decisions about the availability of each channel at each SAP. 
\item \textbf{Proposed multiband}: We implement Alg. \ref{alg:diffusion}, yet assume all SAPs sense the entire spectrum
\item \textbf{Proposed single-band}: We implement Alg. \ref{alg:diffusion}, where each SAP senses a single channel that is assigned via Alg. \ref{alg:scheduler}.
\item \textbf{Centralized}: We consider a core network collecting all energy measurements from all SAPs, and combining them via \emph{equal gain combining} to make a global decision about the availability of each channel. 
\item \textbf{Non-cooperative multiband}: Each SAP senses the entire spectrum and makes a local decision about the availability of each channel. 
\item \textbf{Non-cooperative single-band}: Each SAP randomly picks a channel to sense. The SAP will not have information about other channels. 
\end{itemize}  
We compare the aforementioned schemes in terms of:	
\begin{itemize}
	\item \textbf{Utilization ratio}: The ratio of spatio-spectral blocks that are correctly identified as available by the scheme relative to those found by the genie scheme.
	\item \textbf{Misdetection probability}: The probability of incorrectly deciding a spatio-spectral block is available.
	\item \textbf{Correct decisions}:   The percentage of correct decisions about the channel occupancy, whether available or busy.
\end{itemize}
\noindent We note that the spatio-spectral block is available at the SAP if the energy level of the channel measured at the location of the SAP is below a given threshold.

\begin{figure}[t!]
	\centering
	\begin{subfigure}[t]{.4\textwidth}
		\centering
		\includegraphics[width=2.75in]{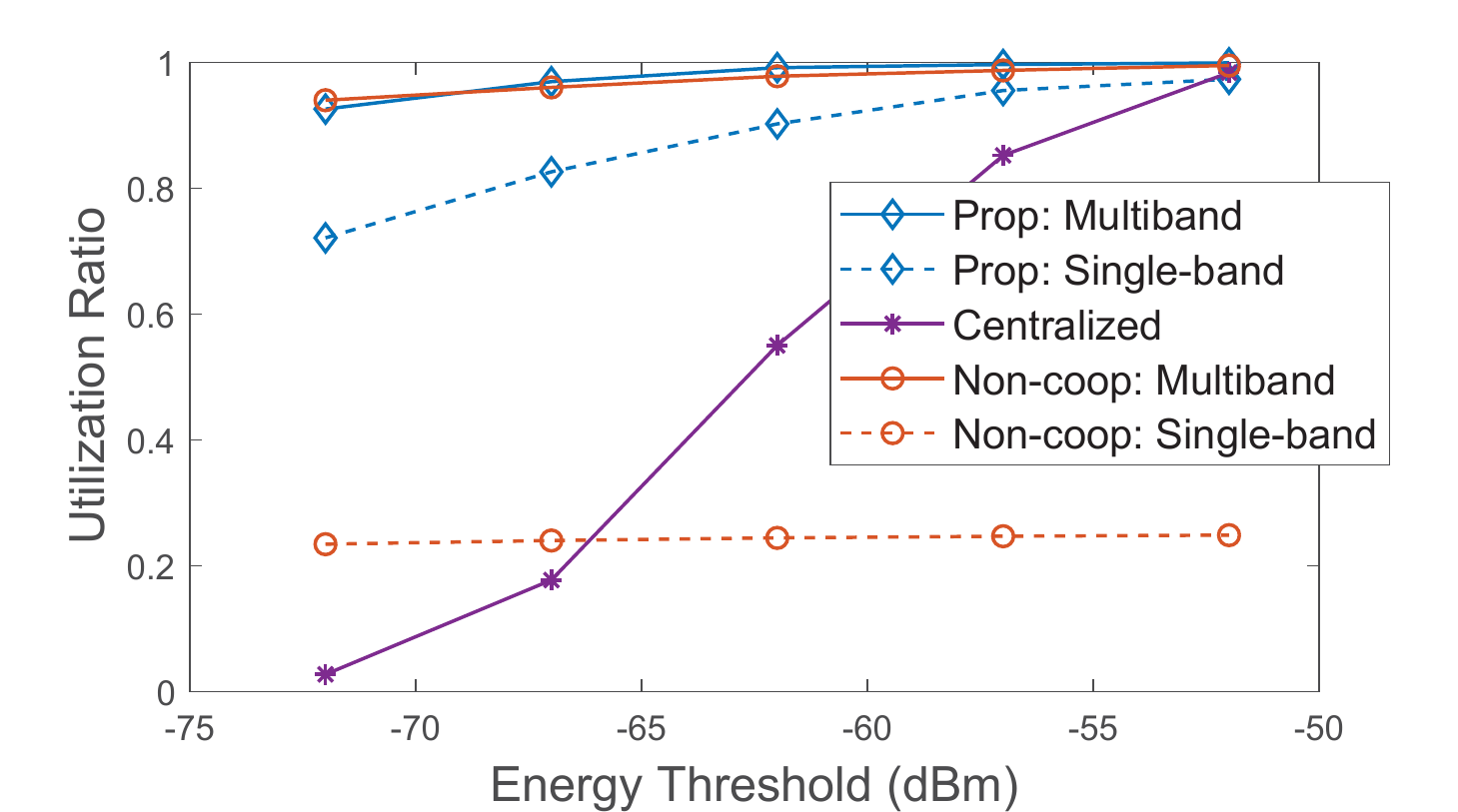}
		\caption{Utilization ratio}
		\label{fig:UT_vs_ED}
	\end{subfigure}\\
	\begin{subfigure}[t]{.4\textwidth}
		\centering
		\includegraphics[width=2.75in]{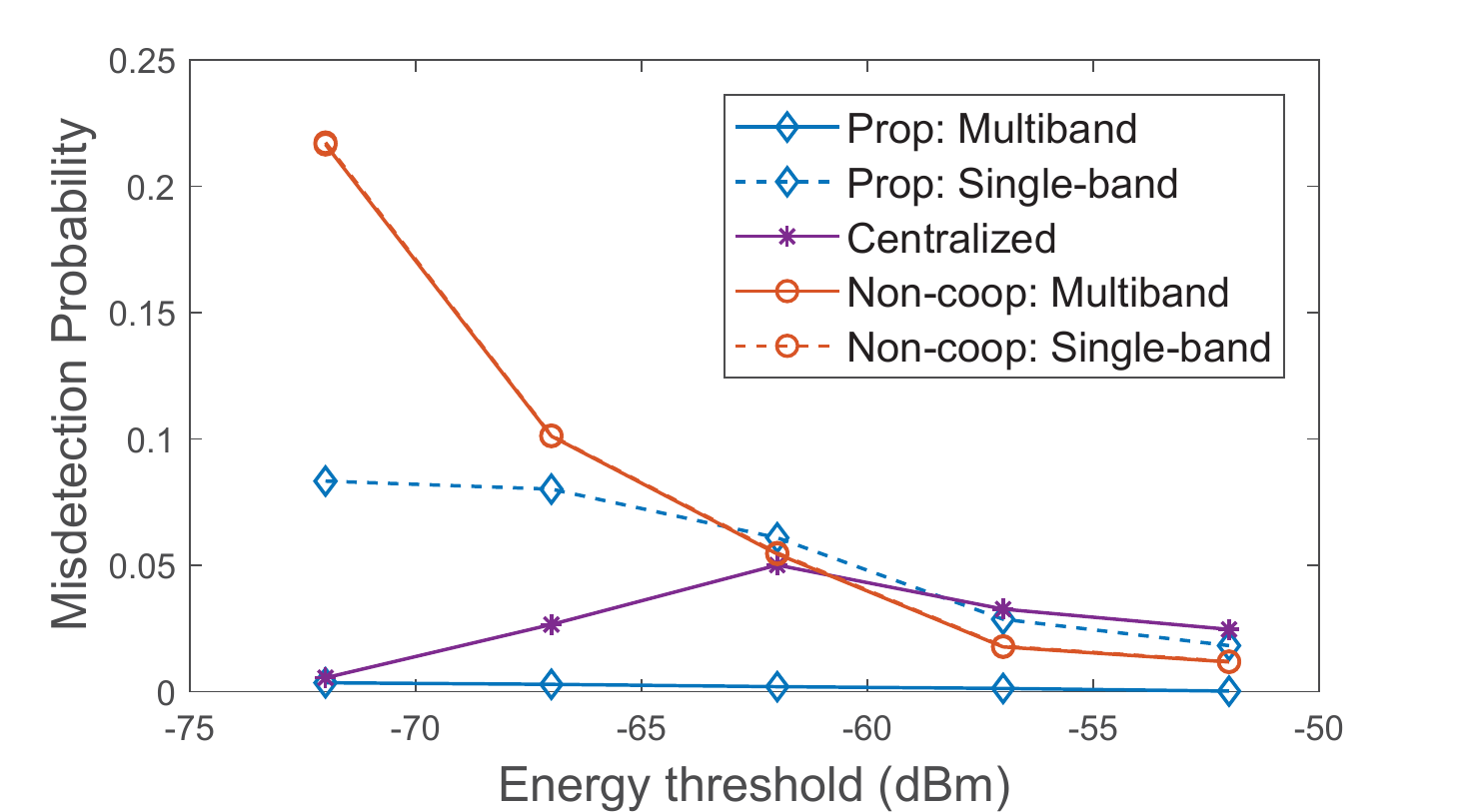}
		\caption{Misdetection probability}
		\label{fig:MD_vs_ED}
	\end{subfigure}
	\caption{Performance with variations of the energy threshold.} 
	\vspace{-.07in}
\end{figure}

Fig. \ref{fig:UT_vs_ED} shows the utilization ratio with variations of the energy thresholds. Clearly, increasing the energy threshold relaxes the coexistence requirement of the network and WiFi APs, and thus more resources can be reused over space and frequency. Comparing the different schemes, we make the following observations. First, both the proposed and non-cooperative multiband solutions identify the highest number of available resources, as each SAP senses the entire spectrum at its location. However, the proposed solution significantly outperforms the non-cooperative one in terms of misdetection, as shown in Fig. \ref{fig:MD_vs_ED}, since the latter may incorrectly decide a busy channel to be available in the presence of a fading channel and/or shadowing. This is not the case with the proposed diffusion algorithm as cooperation helps enhance the reliability of decisions and reduce the misdetection probability. Second, the proposed single-band solution significantly outperforms the non-cooperative single-band scheme although both schemes enforce each SAP to sense a single channel. This follows because by the end of the diffusion-based sensing procedure, each SAP will have occupancy information across all channels, whereas in the non-cooperative one each SAP will be limited to the availability of the sensed channel. Finally, the centralized solution is inefficient for low energy thresholds as a single global decision is made across all SAPs. Thus, if a single SAP has a high energy measurement, it can bias the decision to declare that the channel is busy across all locations, and while misdetection is low for low thresholds, as shown in Fig. \ref{fig:MD_vs_ED}, this comes at the expense of limiting channel reuse over space.  

\begin{figure*}[t!]
	\centering
	\begin{subfigure}[t]{.32\textwidth}
		\centering
		\includegraphics[width=2.25in]{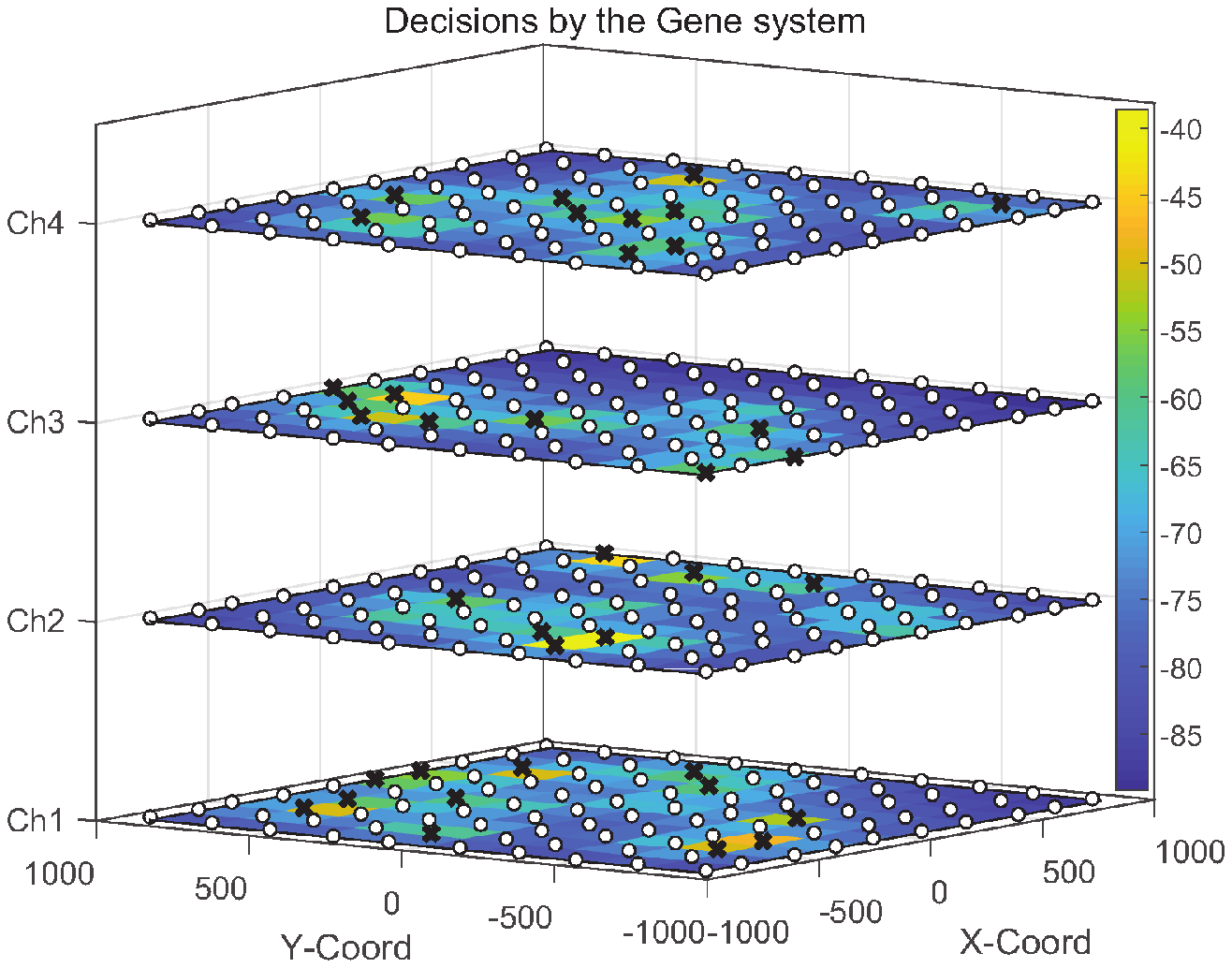}
		\caption{Genie system}
		\label{fig:footprint_genie}
	\end{subfigure}~
	\begin{subfigure}[t]{.32\textwidth}
		\centering
		\includegraphics[width=2.25in]{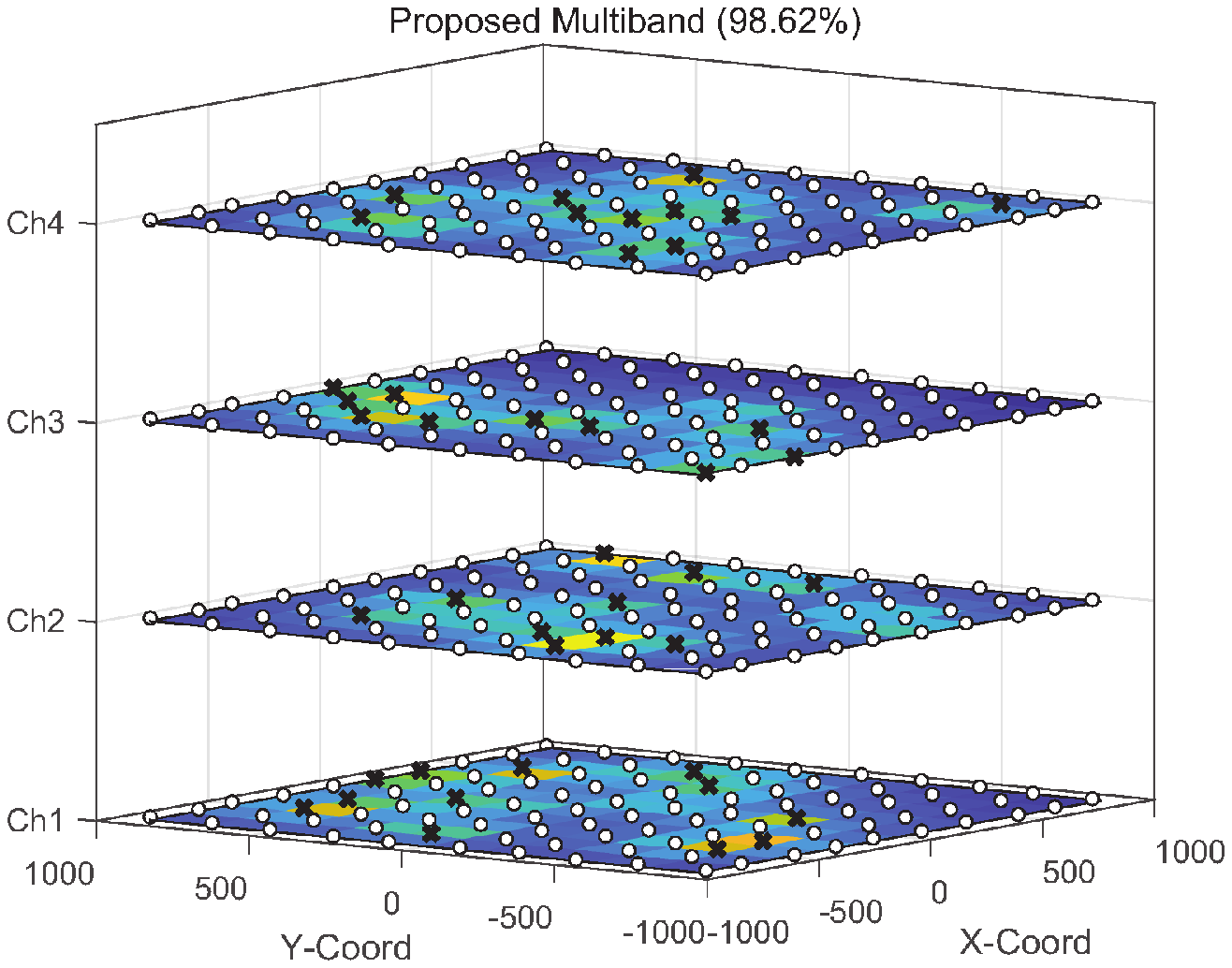}
		\caption{Proposed multiband}
		\label{fig:footprint_propMB}
	\end{subfigure}~
	\begin{subfigure}[t]{.32\textwidth}
		\centering
		\includegraphics[width=2.25in]{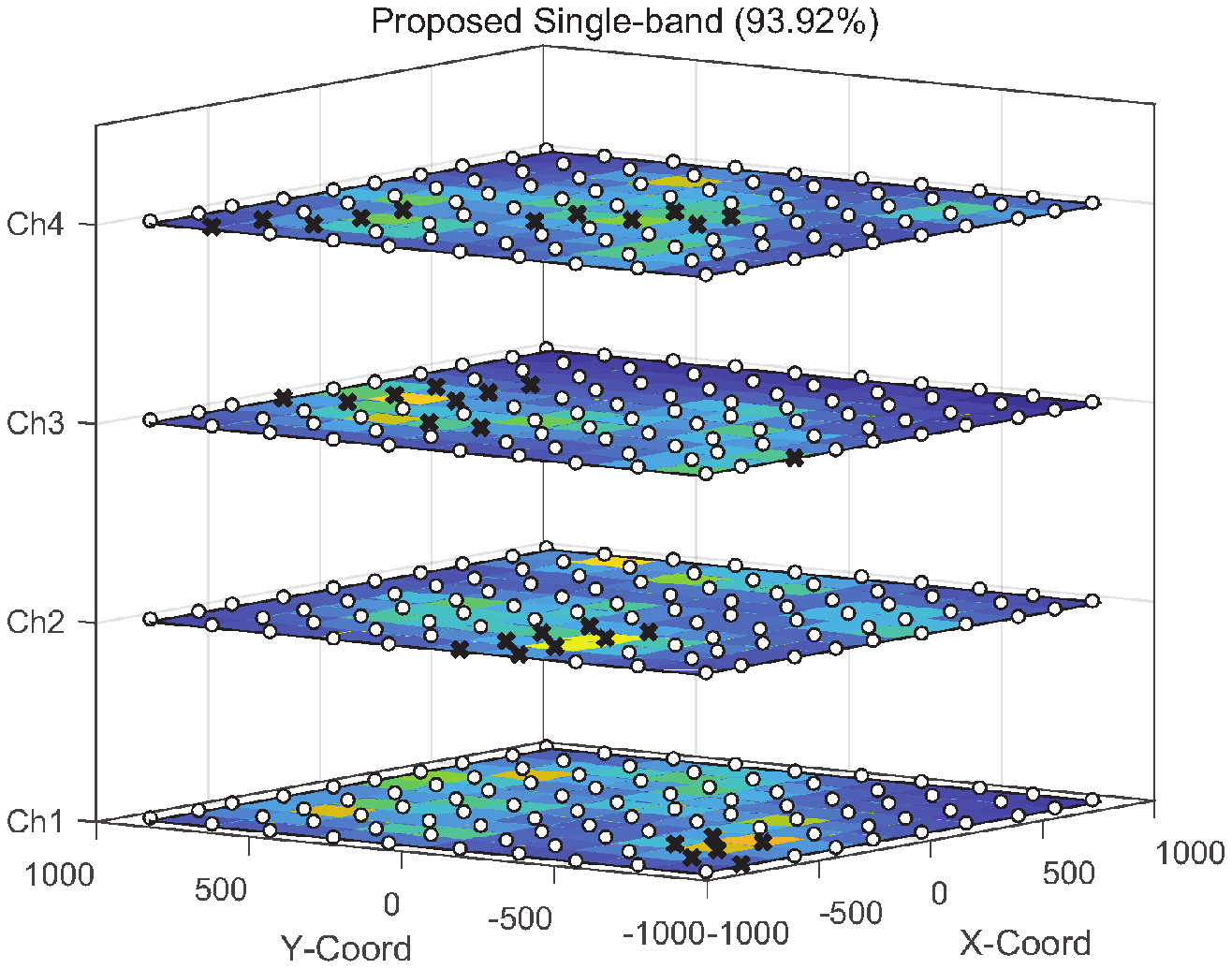}
		\caption{Proposed single-band}
		\label{fig:footprint_propSB}
	\end{subfigure}\\
	\begin{subfigure}[t]{.32\textwidth}
	\centering
	\includegraphics[width=2.25in]{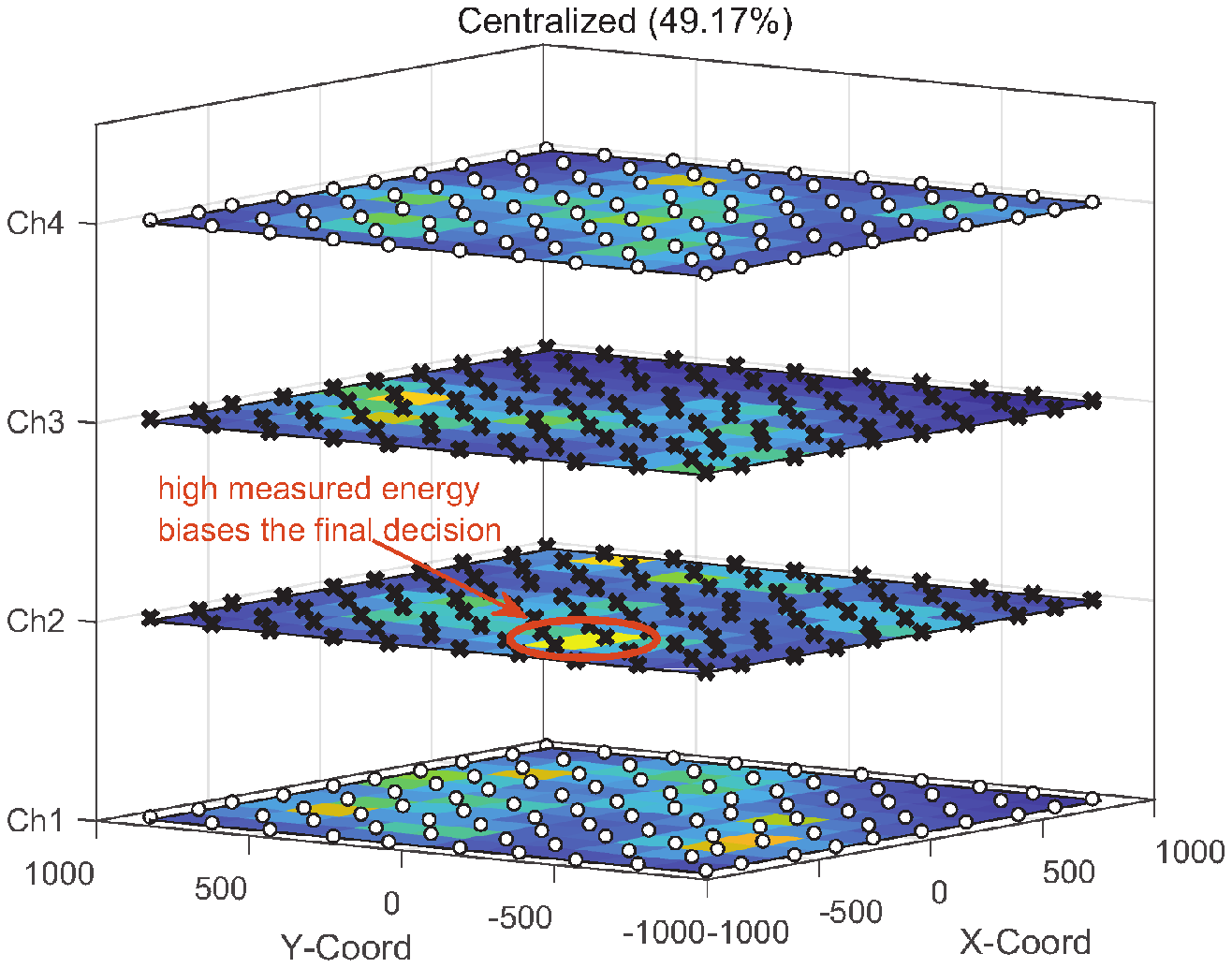}
	\caption{Centralized}
	\label{fig:footprint_centralized}
\end{subfigure}~
\begin{subfigure}[t]{.32\textwidth}
	\centering
	\includegraphics[width=2.25in]{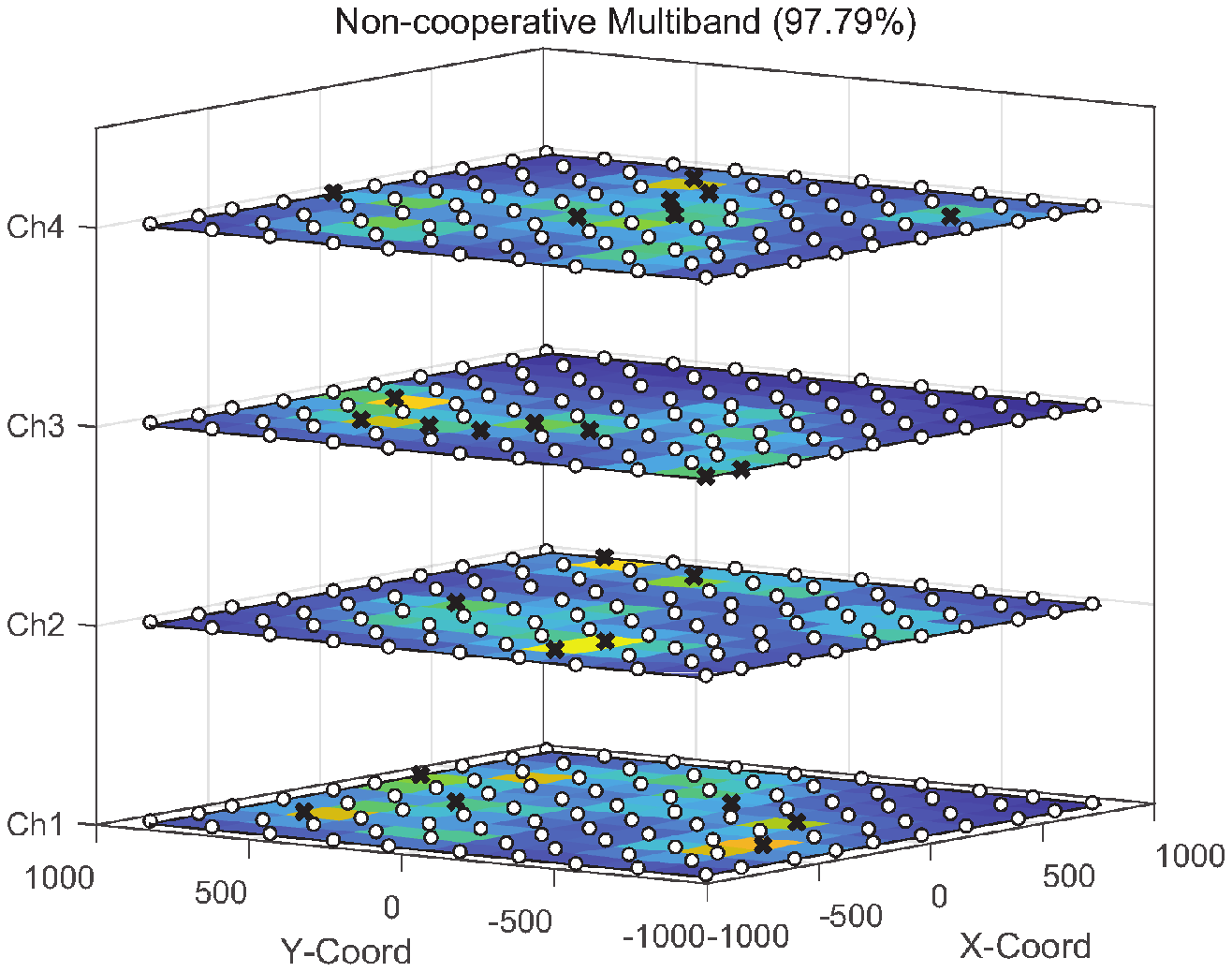}
	\caption{Non-cooperative multiband}
	\label{fig:footprint_nocoopMB}
\end{subfigure}~
\begin{subfigure}[t]{.32\textwidth}
	\centering
	\includegraphics[width=2.25in]{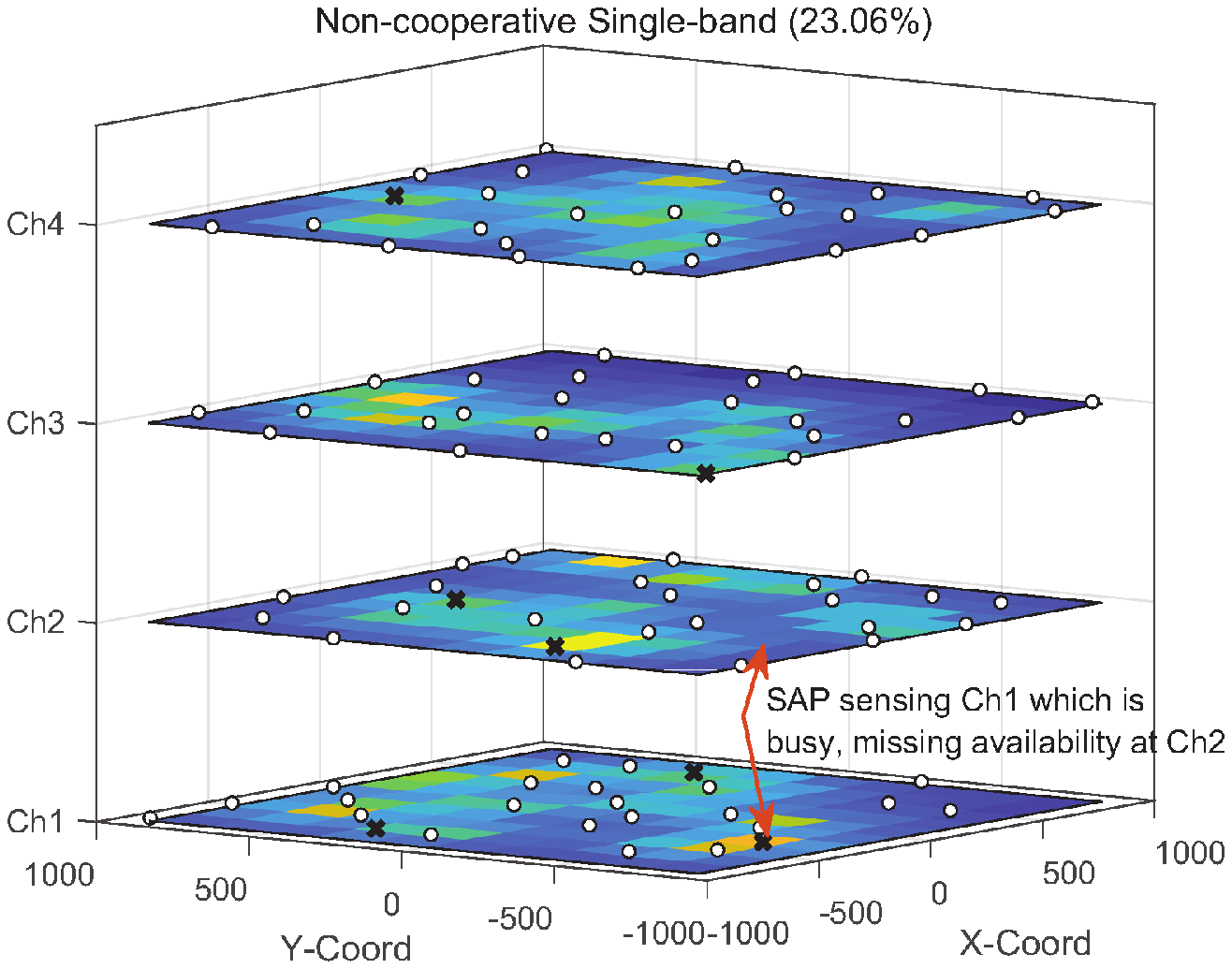}
	\caption{Non-cooperative single-band}
	\label{fig:footprint_nocoopSB}
\end{subfigure}
	\caption{The spatial footprint, i.e., energy measurements, at all SAPs and channels. `$\circ$' and `$\times$' denote available and busy decisions at their locations, respectively. We show in parentheses the percentage of correctly identified spatio-spectral blocks.} 
	\label{fig:footprints}
		\vspace{-.1in}
\end{figure*}

We show in Fig. \ref{fig:footprints} a single snapshot of the spatial energy footprints across all channels, and the corresponding decisions at each SAP for every scheme during that realization. We also show the percentage of correctly identified spatio-spectral blocks in that specific snapshot, where we assume an energy threshold of $-62$dBm. Compared to the genie scheme, the multiband solutions, i.e., the proposed and non-cooperative ones, identify the highest number of available resources, yet the latter scheme achieves this at the expense of high misdetection (cf. Fig. \ref{fig:MD_vs_ED}). The proposed single-band scheme identifies the majority of available resources, albeit at a lower utilization to the proposed multiband scheme since each channel is sensed by 25 SAPs instead of a 100. Note further that in this case, a SAP will rely on nearby SAPs to infer the occupancy of the three channels not sensed by it. Finally, in the centralized processing, it is shown that a single SAP with high energy measurement in Ch2 has biased the final decision to make the channel unavailable in the entire region. 

\begin{figure}[t!]
	\center
	\includegraphics[width=2.75in]{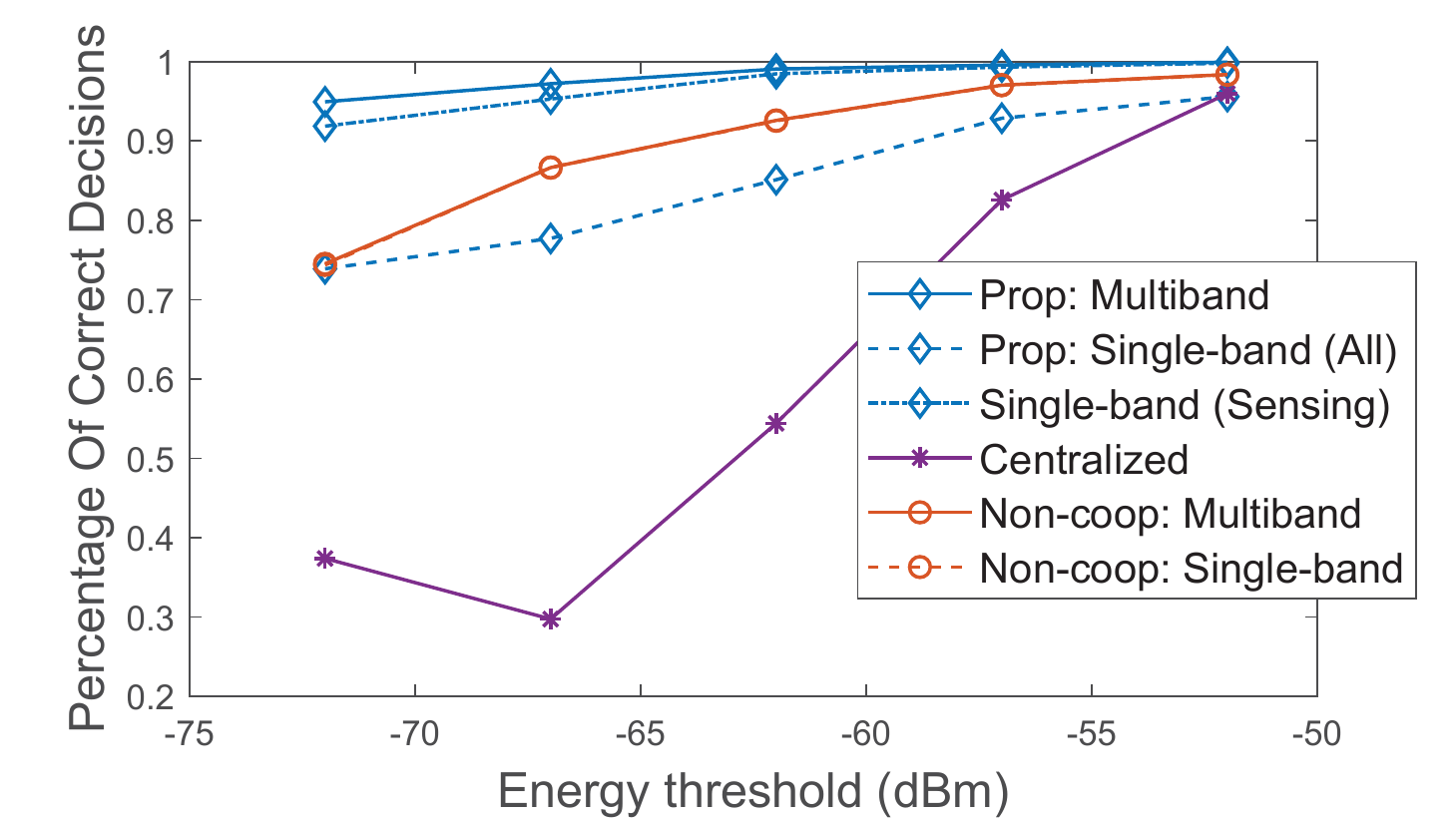}
	\caption{Percentage of correct decisions with variations of the energy threshold.}
	\label{fig:CD_vs_ED}
	\vspace{-0.1in}
\end{figure}

Fig. \ref{fig:CD_vs_ED} shows the percentage of correct decisions made by each scheme. For the proposed single-band solution, we show two curves: (i) one averaging across all SAPs in the system, and (ii) one averaging over the channel decisions made by the SAPs assigned to sense that channel. It is shown that the performance improves for higher energy thresholds as using a higher threshold requires signals to be stronger to be detected at each SAP. We observe that the proposed multiband diffusion outperforms the non-cooperative one as the former has lower misdetection. While the single-band diffusion has fewer correct decisions, across all SAPs, relative to the non-cooperative single-band scheme, the former approximately quadruples the spatio-spectral resource as shown in Fig. \ref{fig:UT_vs_ED}. Note further that looking at SAPs' decisions regarding their sensed channels, the proposed scheme outperforms the non-cooperative one as cooperation improves reliability in the presence of channel impairments, particularly for low energy thresholds. 

\begin{figure}[t!]
	\center
	\includegraphics[width=2.6in]{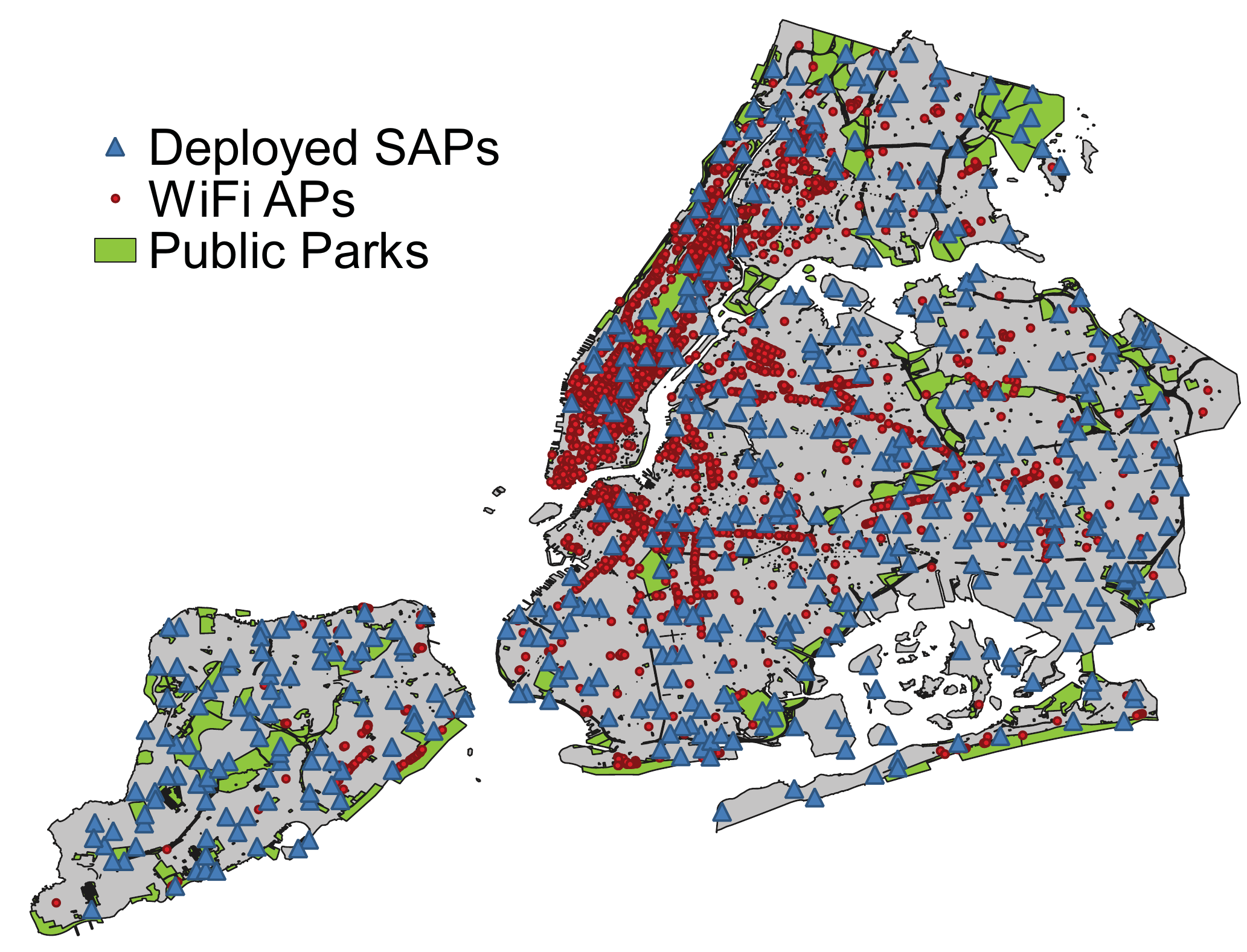}
	\caption{An illustration of the large-scale network set-up.}
	\label{fig:NYC}
\end{figure}

Next, we test the proposed system in a more realistic setting, where we consider a massive IoT application for the public parks in New York City, where sensors and machines can be deployed for water management, tracking traffic activities in the parks, etc. We randomly drop 500 SAPs across the city and deploy 2000 outdoor public WiFi APs, where we use \emph{NYC Open Data} to use their exact coordinates. We treat these APs as interfering transmitters that use the same spectrum. In particular, we consider a 500MHz spectrum centered around 5.43GHz, and the spectrum is channelized into 180KHz channels for NB-IoT operation and 1.4MHz channels for LTE-M operation. Each WiFi AP is assumed to transmit a signal at 30dBm over the signal bandwidth, which is either 20MHz, 40MHz, or 80MHz within the wideband spectrum. We compare the proposed distributed sensing algorithm with non-cooperative multiband and non-cooperative single-band. In the proposed system, the assignment scheduler assigns each SAP a subset of channels with a total bandwidth of 20MHz. Finally, we randomly drop $1\times 10^5$ IoT objects in the public parks. An illustration of the network is given in Fig. \ref{fig:NYC}.  

\begin{figure}[t!]
	\center
	\includegraphics[width=2.75in]{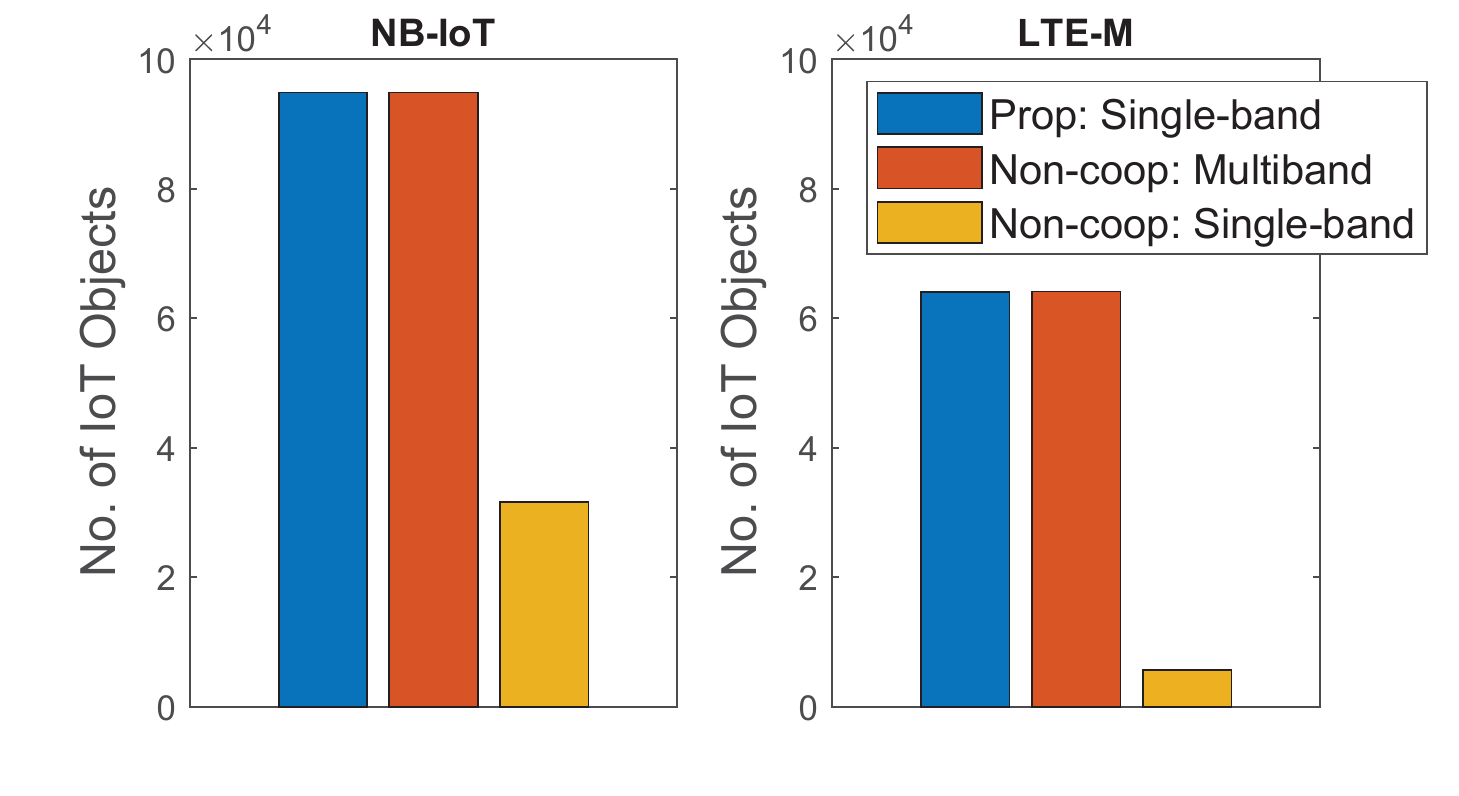}
	\caption{Average number of scheduled devices per scheme.}
	\label{fig:CaseStudySim}
	\vspace{-0.1in}
\end{figure}

Fig. \ref{fig:CaseStudySim} shows the average number of devices that are scheduled in the system over correctly identified available channels for each scheme. It is observed that for NB-IoT, almost all devices are scheduled via the proposed and the non-cooperative multiband schemes. It is also clear that the proposed system significantly outperforms the non-cooperative single band-solution, emphasizing that cooperation across neighboring SAPs is not only beneficial to enhance the reliability of sensing a particular channel, but also useful to infer the occupancy of other channels, given a proper sensing assignment. Similar trends hold for LTE-M, yet fewer devices are scheduled in comparison to NB-IoT, as each device requires larger bandwidth. 

\section{Conclusions}\label{sec:conclusion}
The unlicensed spectrum access via a dense skeleton of APs with sensing capabilities has the potential to connect a massive number of IoT objects by exploring a large pool of narrowband channels and aggressively reusing them over space. Instead of using wideband sensing at each SAP to explore a wide swath of spectrum, we have proposed a sensing assignment scheduler so that the SAP senses a small subset of the spectrum, reducing the sensing complexity. To obtain local information about the entire spectrum, we have developed a distributed sensing algorithm, which requires SAPs to share measurements only among their neighbors. Simulation results have demonstrated the effectiveness of the proposed system in reliably identifying the available spatio-spectral blocks.

\bibliographystyle{IEEEtran}
\bibliography{C:/Users/ghait/Dropbox/References/IEEEabrv,C:/Users/ghait/Dropbox/References/References}

\end{document}